\documentclass[11pt]{article}
\usepackage{Blois_eprint,epsfig}

\def\be{\begin{equation}}
\def\ee{\end{equation}}
\def\bea{\begin{eqnarray}}
\def\eea{\end{eqnarray}}

\begin{document}
\input{psfig.sty}
\vspace*{2cm}
\begin{center}
\Large{\textbf{XIth International Conference on\\ Elastic and Diffractive Scattering\\ Ch\^{a}teau de Blois, France, May 15 - 20, 2005}}
\end{center}

\vspace*{2cm}
\title{TWENTY YEARS OF DIFFRACTION AT THE TEVATRON}

\author{ K. GOULIANOS }

\address{The Rockefeller University, 1230 York Avenue, New York, NY 10021, U.S.A}

\maketitle\abstracts{
Results on diffractive particle interactions from the Fermilab Tevatron 
$\bar pp$ collider are placed in perspective  
through a QCD inspired  phenomenological 
approach, which  exploits scaling and factorization 
properties observed in data. The results discussed are those obtained 
by the CDF Collaboration from a comprehensive set of single, double, 
and multigap soft and hard diffraction processes studied during 
the twenty year period since 1985, when the CDF diffractive program was 
proposed and the first Blois Workshop was held.
}
Diffractive $\bar pp$ interactions are characterized by the 
presence of at least one large rapidity gap, defined as a  region 
of pseudorapidity~\footnote{We use rapidity, $y=\frac{1}{2}\frac{E+p_L}{E-p_L}$, and pseudorapidity, $\eta=-\ln\tan\frac{\theta}{2}$, interchangeably, since in the kinematic region of interest the values of these two variables are approximately equal.}
devoid of particles.
A diffractive rapidity gap, which  may 
be forward (adjacent to a leading nucleon) 
or central, is presumed to be formed by the exchange 
of a {\em Pomeron}~\cite{Regge}, which in QCD is a  color singlet 
quark/gluon object with vacuum quantum numbers. 
Diffraction in which there is a high momentum-transfer 
partonic scattering in the event  
is referred to as {\em hard diffraction}~\cite{ingeschlein}$^,$~\footnote{
G.~Ingelman, ``Hard Diffraction - from Blois 1985 to 2005,'' 
in these Proceedings.}. 
In this paper, 
%
%
we briefly review what we have learned about diffraction 
from the Collider Detector at 
Fermilab (CDF) in Run-I (1989-1995)  of the 
Tevatron $\bar pp$ collider operating at 1800 GeV c.m.s. energy,  
and comment on the goals, results already obtained~\cite{dis05_CDF}$^,$~\footnote{See also C. Mesropian, 
``New Diffraction Results from CDF,'' in these Proceedings.}, and expectations 
from the 1960 GeV Run-II program,  
which is currently under way. 
The CDF results are placed 
in perspective using a QCD based phenomenological 
approach, which exploits scaling and factorization properties observed in the 
data (see Ref.~\cite{lathuile}).
\section{Run-I Results}
In addition to measuring $\bar pp$ elastic, single diffraction (SD), 
and total cross sections at $\sqrt s=540$ and 1800 GeV, CDF studied
several soft and hard diffraction processes at $\sqrt s=1800$ GeV, 
and in some cases at $\sqrt s =630$ GeV~\cite{lathuile}. 
Soft processes studied include:
\begin{center}
\begin{tabular}{lll}
{\bf DD}&Double Diffraction&$\bar{p}p\rightarrow X+{\rm gap}+Y$\\
{\bf DPE}&Double Pomeron Exchange&$\bar{p}p\rightarrow \bar{p}+{\rm gap}+X+{\rm gap}+p$\\
{\bf SDD}&Single $\oplus$ Double Diffraction&
$\bar{p}p\rightarrow \bar{p}+{\rm gap}+X+{\rm gap}+Y$\\
\end{tabular}
\end{center}
In the area of hard diffraction, CDF measured SD dijet, $W$, $b$-quark and
$J/\psi$ production, DD dijet production, 
and DPE dijet production.
Schematic diagrams and event topologies for representative  
diffractive processes studied in Run-I are shown in Fig.~\ref{fig:diagrams}.
 
\begin{minipage}[t]{0.5\textwidth}
\phantom{xxx}
\centerline{\bf\it\sc SOFT DIFFRACTION}
\vglue 1em
\hspace{-0.5em}\includegraphics[width=0.95\textwidth]{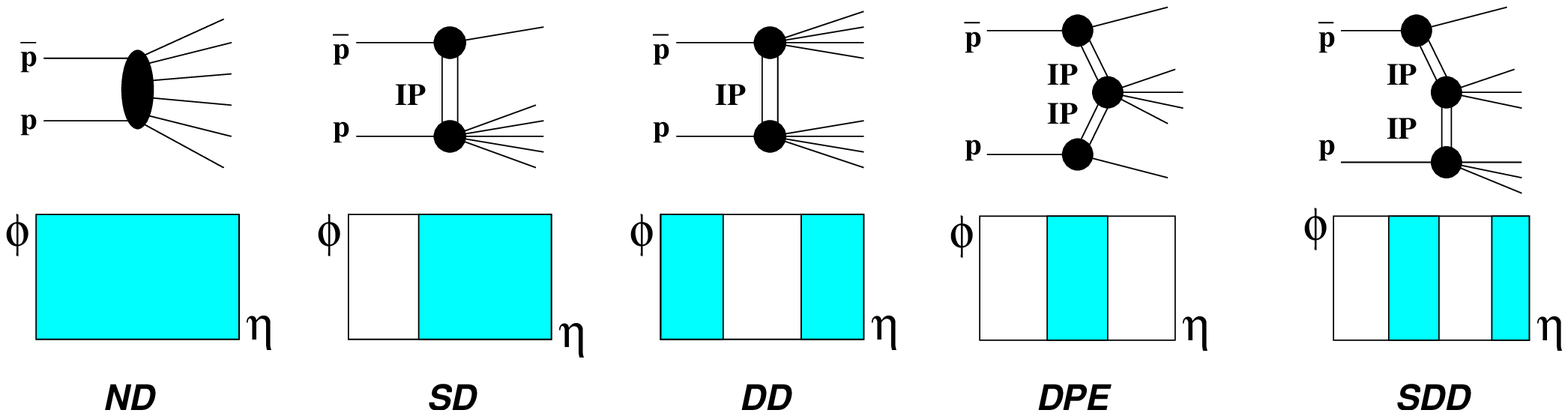}
\end{minipage}
\hspace{0.02\textwidth}
\begin{minipage}[t]{0.5\textwidth}
\phantom{xxx}
\centerline{\bf\it\sc HARD DIFFRACTION}
\vspace{1em}\vspace{-0.5ex}
\hspace{-1em}\includegraphics[width=1.1\textwidth]{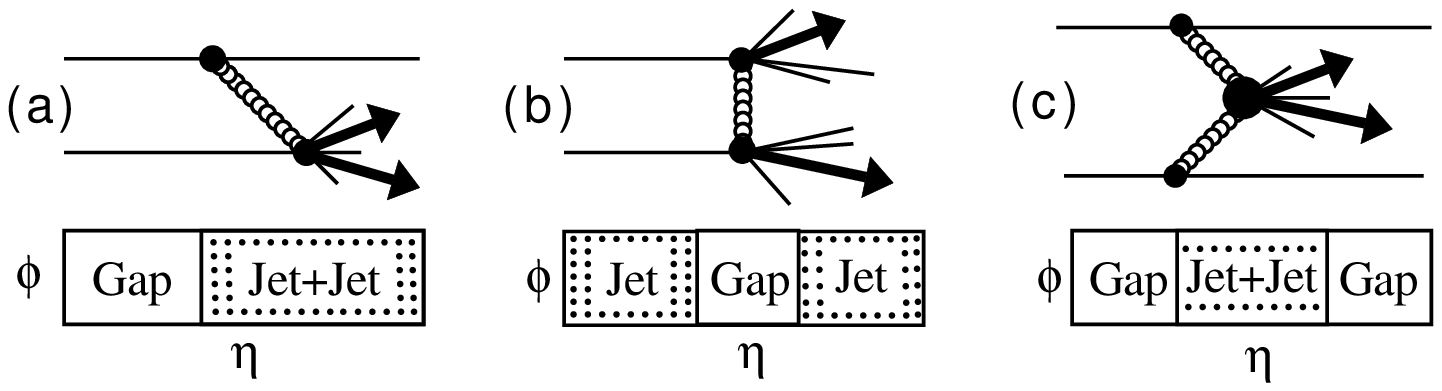}
\vspace{-22em}
\end{minipage}
\vspace{-5em}
\begin{figure}[h]
\caption{Schematic diagrams and $\eta$-$\phi$ topologies of representative 
diffractive processes studied by CDF in Run-I. 
The shaded areas represent regions of 
pseudorapidity in which {\em there is} particle production.}
\label{fig:diagrams}
\end{figure}

Two types of hard diffraction results were obtained in Run-I: 
diffractive to non-diffractive cross 
section ratios, using the rapidity gap signature 
to select diffractive events, 
and diffractive to non-diffractive structure function ratios,  using a 
Roman Pot Spectrometer (RPS) to trigger on leading 
antiprotons~(see Fig.~\ref{fig:detectors},~{\em{left}}). 

\begin{minipage}[t]{0.5\textwidth}
\phantom{xxx}
\vglue -1em
\hspace{-0.2\textwidth}\psfig{figure=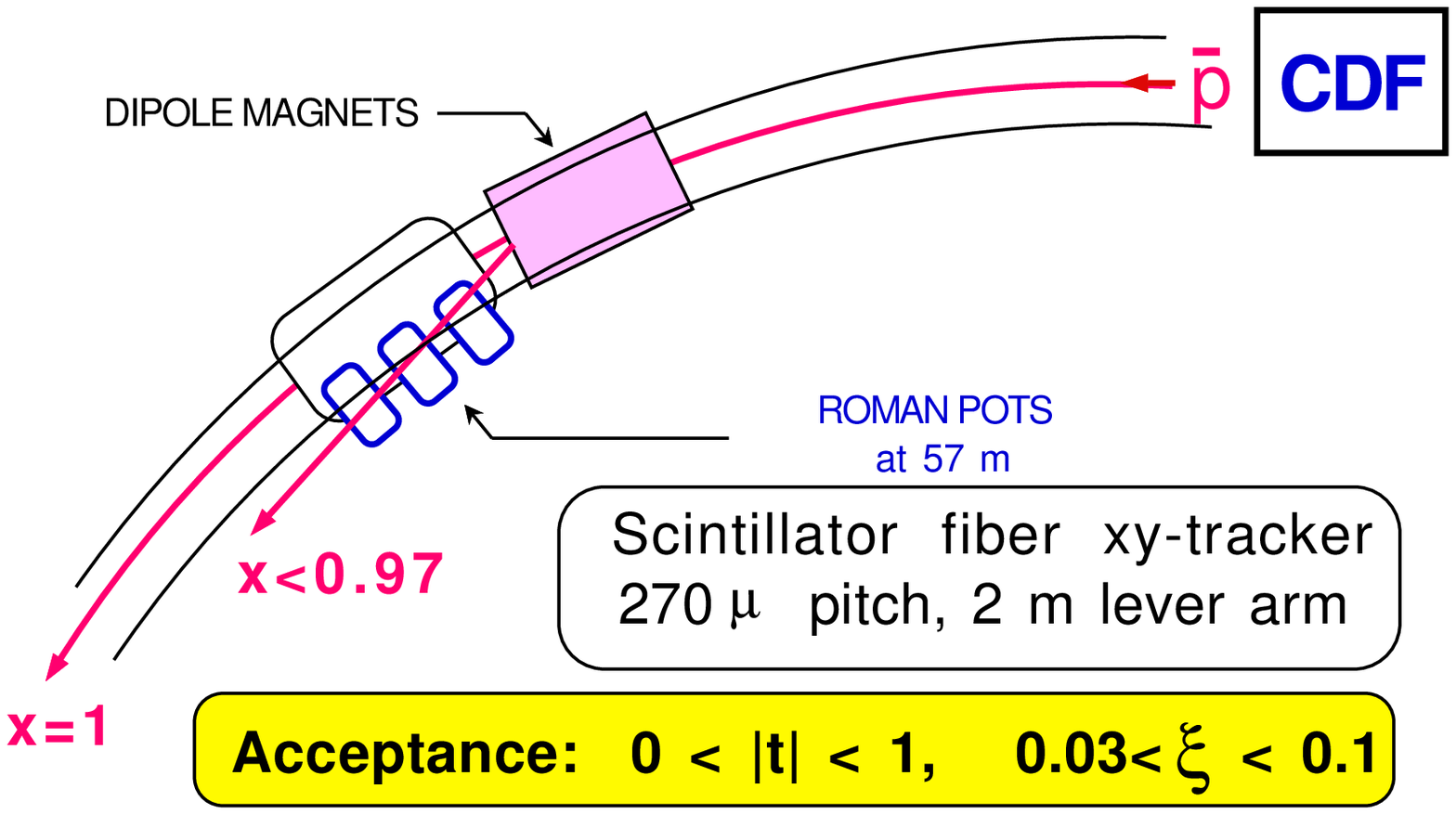,width=1.2\textwidth}
\vspace{-21em}
\end{minipage}
\hspace{-3em}
\begin{minipage}[t]{0.55\textwidth}
\phantom{xxx}
\vspace{-2em}
\psfig{figure=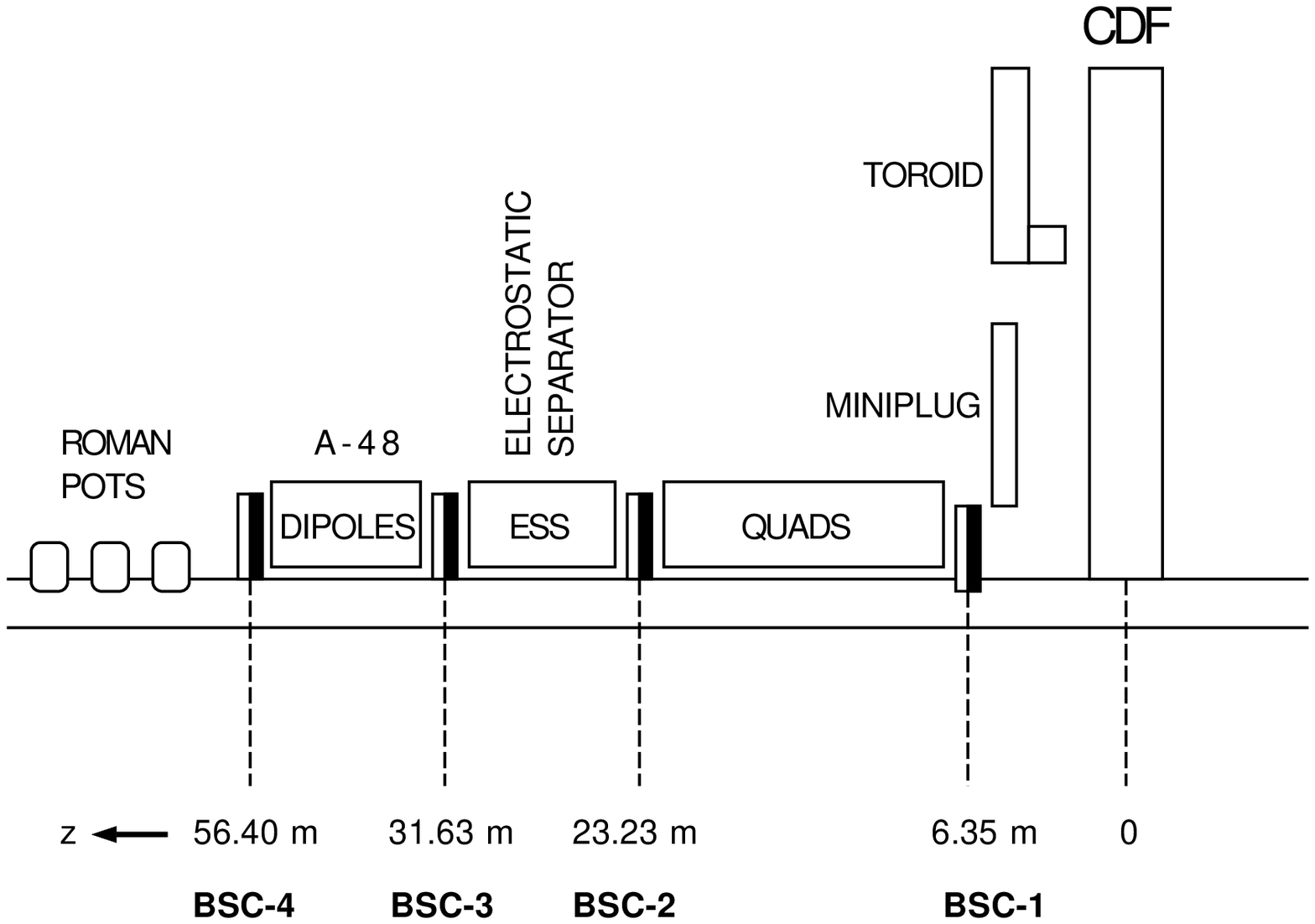,width=1\textwidth}
\end{minipage}
\vspace*{-15em}
\begin{figure}[h]
\caption{Layout of the CDF detector in Run-I (\em{left}) and in Run-II (\em{right}), showing the special forward detectors used in the 
diffractive program.}
\label{fig:detectors}
\end{figure}

The Run-I diffractive production results exhibit regularities in normalization
and factorization properties pointing to
the QCD character of diffraction.
The result that has attracted the most attention is the  
breakdown of QCD factorization, indicated by an 
$\sim{\cal{O}}(10)$ suppression in normalization of the diffractive 
structure function (DSF) measured from diffractive dijet production 
at the Tevatron relative to that measured 
from diffractive deep inelastic scattering (DDIS) at HERA. However, 
less attention has been paid to the remarkable $s$-{\em independence} 
of the $d\sigma^{SD}/dM^2$ diffractive differential cross section, 
a scaling  property that seems to regulate the magnitude of the breakdown 
of factorization. This ``$M^2$-scaling'' behavior has 
profound implications about the mechanism of diffraction,
favoring a composite over a particle-like Pomeron,  
as discussed in Ref.~\cite{lathuile} and in 
original references therein.
    
\subsection{Breakdown of Factorization}
At $\sqrt s=$1800 GeV, the SD/ND ratios (gap fractions)   
for dijet, $W$, $b$-quark, and $J/\psi$ production, as well the ratio of
DD/ND dijet production, are all $\approx 1\%$.
These ratios are suppressed by a factor of $\sim$10 
relative to standard QCD inspired theoretical 
expectations ({\em e.g.} 2-gluon exchange), 
or relative to predictions based on 
diffractive parton densities measured from DDIS at HERA. 
This suppression represents a severe breakdown of QCD 
factorization. 

There are two interesting features characterizing the data: 
(i) despite the overall suppression in normalization, 
factorization approximately holds among different diffractive 
processes at fixed $\sqrt s\,$, and (ii)
the magnitude of the suppression is comparable to that
observed in soft diffraction processes relative to Regge theory expectations.
These features indicate that the suppression is in the formation of the 
rapidity gap. The generalized gap renormalization model  provides 
a good description of the data (see Ref.~\cite{lathuile}). 

\subsection{Restoration of Factorization in Multigap Diffraction}
Another interesting aspect of the Run-I results is that 
ratios of two-gap to one-gap cross sections 
for both soft and hard processes appear to obey factorization. This feature 
of the data provides both a clue to understanding diffraction and 
a tool for diffractive studies using processes with 
multiple rapidity gaps (see Refs.~\cite{lathuile}$^,$\cite{DIS05_multigap}).

\section{Run-II Program}
New diffractive results from CDF obtained in Run-II are presented in 
these Proceedings by C.~Mesropian, but are included here for completeness.
The Run-II diffractive program of CDF 
(see Fig.~\ref{fig:detectors},~{\em right}) 
is aiming at deciphering the QCD nature 
of the Pomeron by measuring the dependence of the diffractive 
structure function on $Q^2$, $x_{Bj}$, $t$, 
and $\xi$ (fractional momentum loss of the diffracted nucleon) 
for different diffractive production processes; in addition, 
the possibility of a composite Pomeron is being investigated by studies of 
very forward jets with a rapidity gap between jets.
Another goal of the program is to measure exclusive production rates 
(dijet, $\chi_c^0$, $\gamma\gamma$), 
which could be used to establish 
benchmark calibrations for 
exclusive Higgs production at LHC~\cite{KMR,KKMRS}.  
Preliminary results from data collected at $\sqrt s=1960$ GeV  
confirm the Run-I DSF results~\cite{lathuile,MGmoriond}.
New results from Run-II are the measurement of the 
$Q^2$ dependence of the DSF obtained 
from dijet production and 
limits on exclusive production rates. 
\vglue -0.5em
\hspace*{-1em}\begin{minipage}[t]{0.5\textwidth}
\phantom{xxx}
\hspace{-1em}\includegraphics[width=1.05\textwidth]{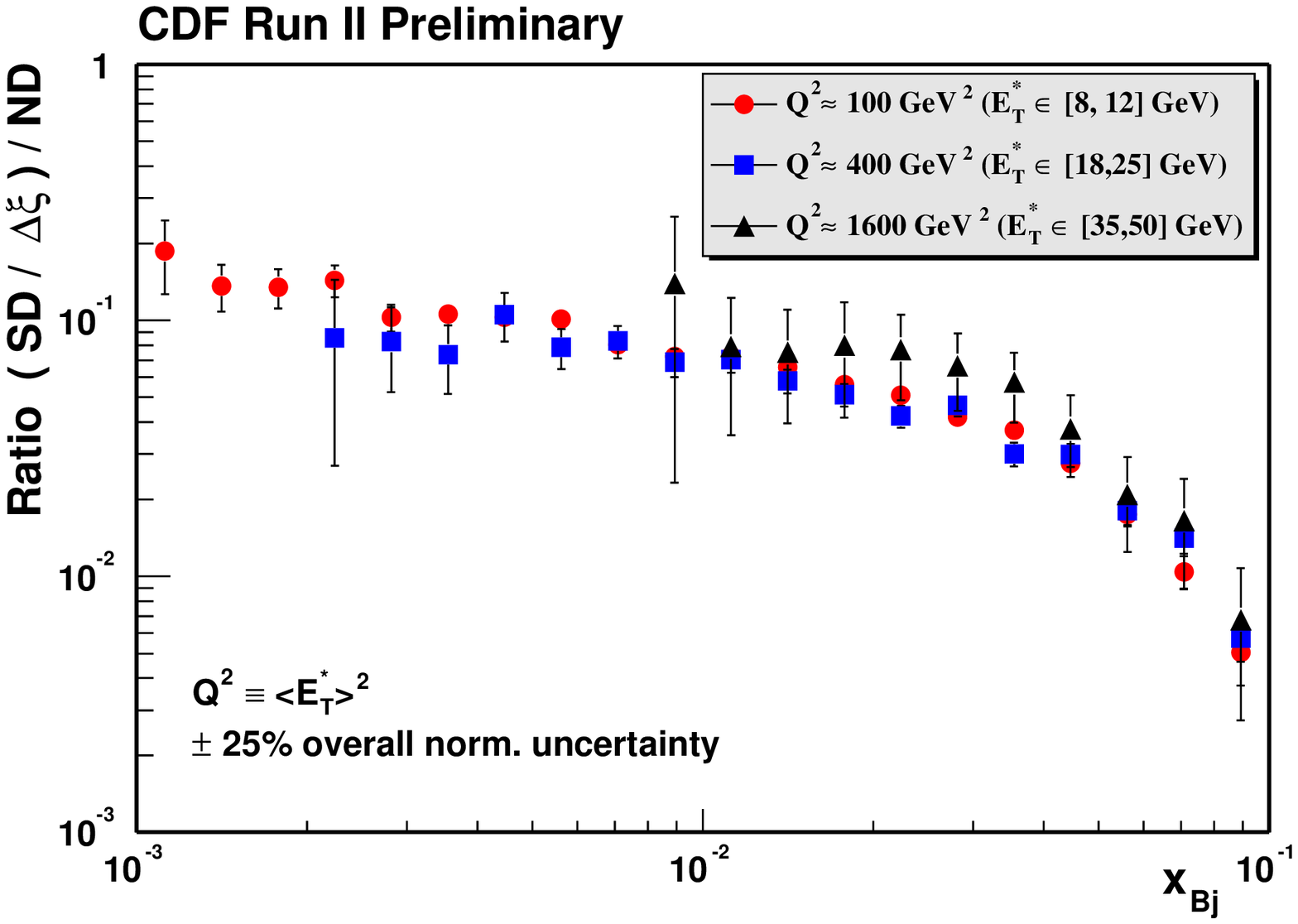}
\end{minipage}
\hspace{-2ex}
\begin{minipage}[t]{0.5\textwidth}
\phantom{xxx}
\hspace{-1em}\includegraphics[width=1.14\textwidth]{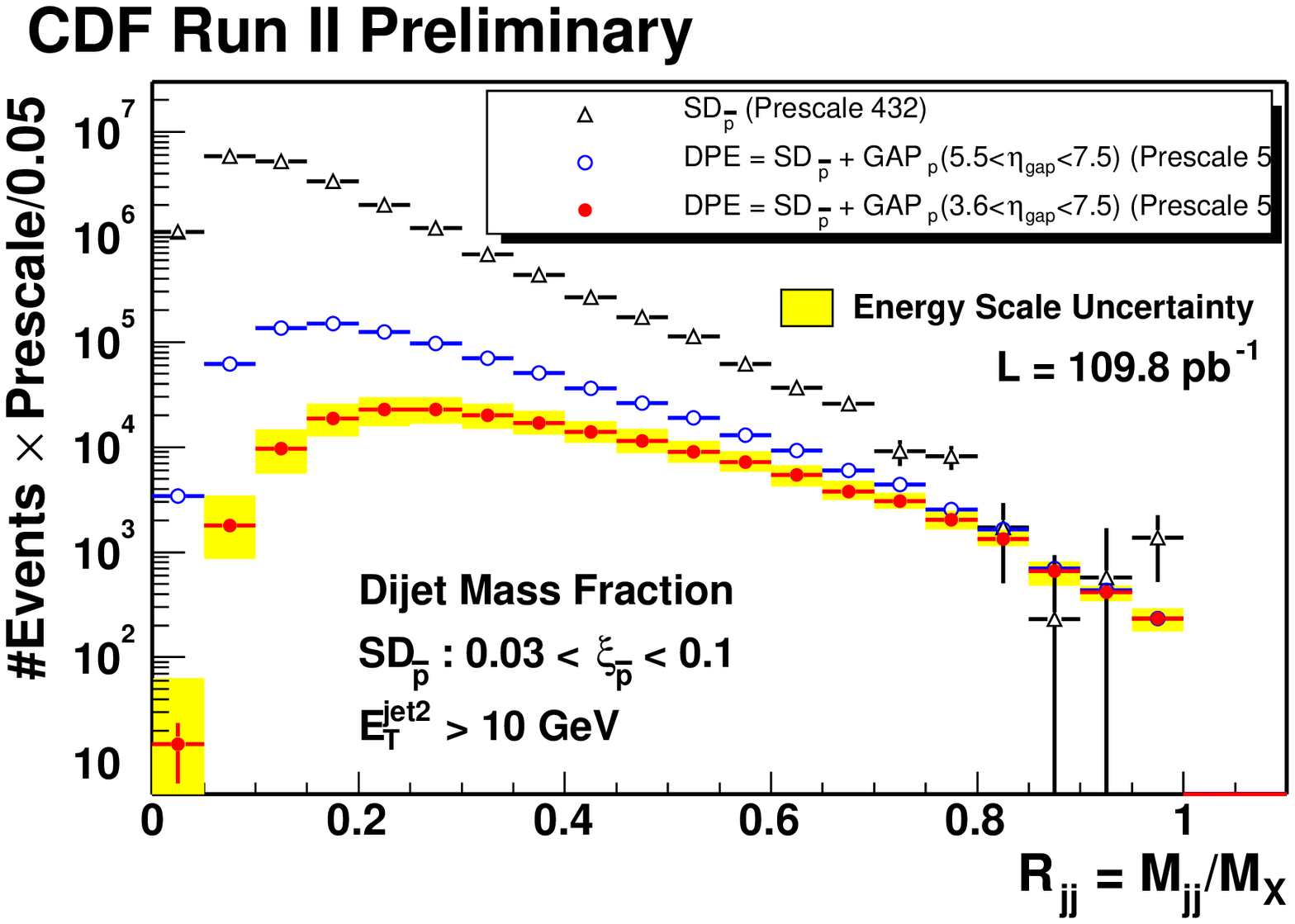}
\end{minipage}

\vglue -4ex
\begin{figure}[ht]
\caption{({\em left}) Ratio of SD/$\Delta\xi_{\bar{p}}$
over ND rates  obtained
from dijet data at various $Q^2$ ranges; {(\em right})  
ratio of dijet mass to 
total mass ``visible'' in the calorimeters for dijet production in events 
with a leading antiproton within $0.3<\xi_{\bar{p}}<0.1$ and various 
gap requirements on the proton side: ({triangles}) no gap requirement, 
({open circles}) gap in 
$5.5<\eta<7.5$, and ({filled circles}) gap in region $3.5<\eta<7.5$.}
\label{fig:Q2Mjj}
\end{figure}
 
\subsection{The Diffractive Structure Function}
In Fig.~\ref{fig:Q2Mjj}~{(\em left)},
the ratio of SD/ND rates, which in LO QCD and at fixed $x_{Bj}$ 
is equal to the ratio of the corresponding structure functions, 
shows no appreciable $Q^2$ dependence. 
This result was foreseen in the renormalization model~\cite{newapproach}, 
in which the diffractive structure function is basically the low-$x$ 
($x<\xi$) structure function of the diffracted nucleon. 
More data are currently being analyzed to improve the statistics 
of this measurement.
Data are at hand and analyses are in progress for the measurement of 
the $t$, $\xi$, and flavor dependence of the DSF 
using dijet, $W$, and $J/\psi$ production. In addition, factorization 
will be tested more accurately than in Run-I by comparing the 
DSFs obtained from dijet production in SD and DPE.    

\subsection{Exclusive Production}
\subsubsection{Exclusive Dijet Production}
The search for exclusive dijet production is based on measuring the 
dijet mass fraction $M_{jj}$, 
defined as the mass of the two leading jets in an event 
divided by the total mass 
reconstructed from all the energy observed in all calorimeters.
Fig.~\ref{fig:Q2Mjj} ({\em right}) shows $M_{jj}$ distributions for events 
with different selection criteria. The signal from exclusive dijets
is expected to be concentrated in the region of $R_{jj}>0.8$,
with values of $R_{jj}<1$ being caused by measurement resolution effects and 
final state radiation. Background events from 
inclusive DPE production, 
$\bar p p\rightarrow (\bar p+gap)+JJ+X+gap$, are expected to contribute to the 
entire $M_{jj}$ region. 

\hspace*{-1.5em}\begin{minipage}[t]{0.5\textwidth}
\phantom{xxx}
\includegraphics[width=1.1\textwidth]{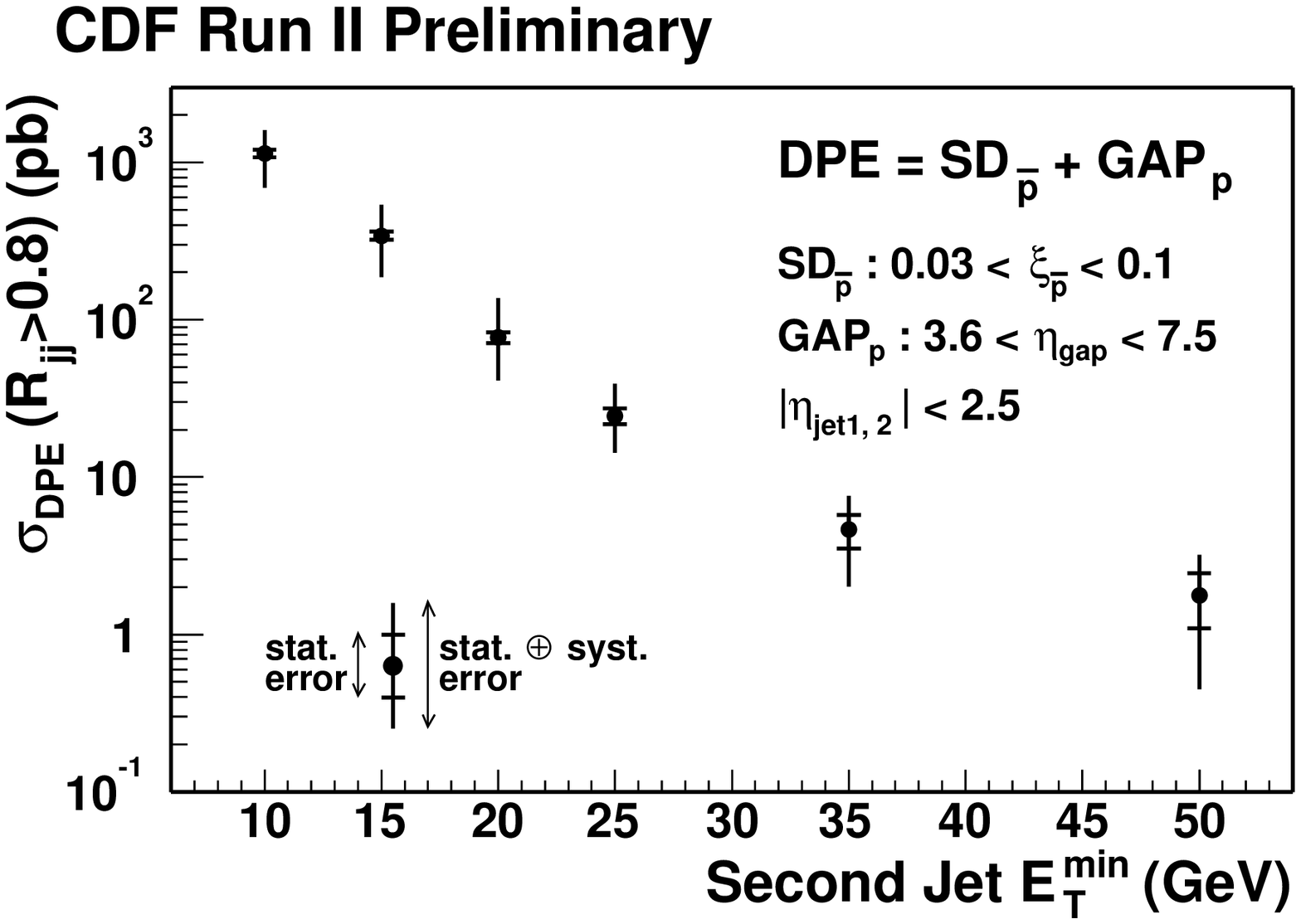}
\end{minipage}
\hspace{1ex}
\begin{minipage}[t]{0.5\textwidth}
\phantom{xxx}
\vspace{-0.3em}
\includegraphics[width=1\textwidth]{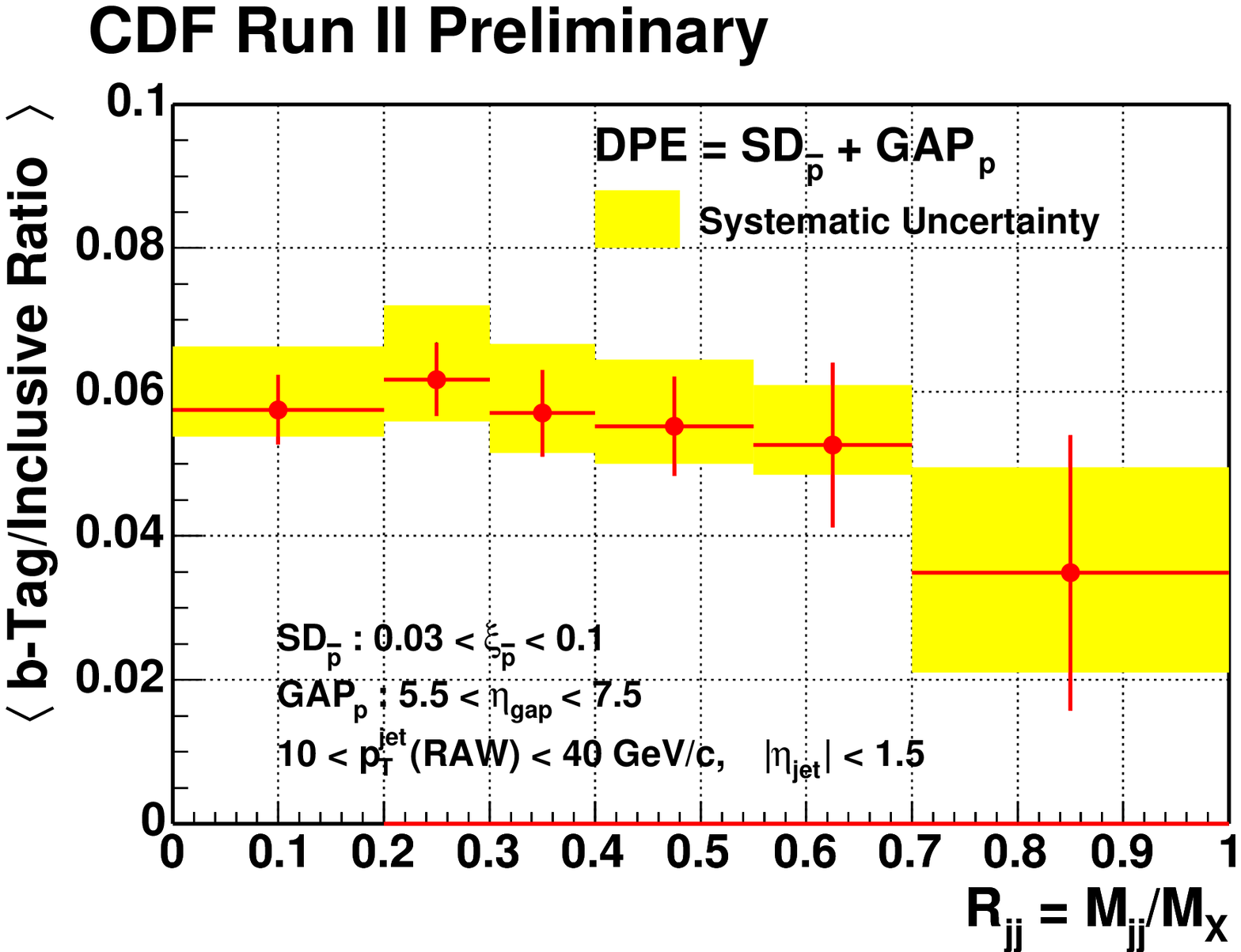}
\end{minipage}

\vglue -1em
\begin{figure}[ht]
\caption{
({\em left}) 
Dijet production cross sections for $R_{jj}>0.8$ in DPE events 
as a function of $E_T^{min}$, the $E_T$ of the next to the highest 
$E_T$ jet; 
({\em right}) the ratio of $b$-tagged to all jets in the DPE 
dijet event sample versus the dijet mass fraction.   
}
\label{fig:exclusiveJJ}
\end{figure}

Since no peak is observed at $R_{jj}>0.8$ in 
Fig.~\ref{fig:Q2Mjj} ({\em right)}, CDF reports production 
cross sections for events with $R_{jj}>0.8$, which could be used 
as upper limits for exclusive production. 
Figure~\ref{fig:exclusiveJJ}~({\em left)} 
shows such cross sections for various kinematic cuts plotted versus 
$E_T^{min}$, the next to leading jet $E_T$. These cross 
sections agree, within errors, with recent predictions for exclusive dijet 
production~\cite{KMR,KKMRS}. Thus, for the predictions to be correct, 
the background would have to vanish as $R_{jj}\rightarrow 1$. While this is 
guaranteed by the $J_z=0$ selection rule for leading order 
$gg\rightarrow q\bar q$ jets of
$m_q<<M_{jet}$, Monte Carlo (MC) simulations are used to deal 
with the dominant $gg\rightarrow gg$ process. 
To avoid using simulations, one could use $q\bar q$ events to estimate
the background. This can be done using dijet events in which at least 
one of the jets is $b$-tagged. 
Figure~\ref{fig:exclusiveJJ}~({\em right}) shows the ratio of 
$b$-tagged to inclusive dijet events versus dijet mass fraction.
A suppression is observed as $M_{jj}\rightarrow 1$, as 
would be expected if there were exclusive dijets in the sample. 
However, background still may exist 
from the gluon splitting process 
$gg\rightarrow g+g(\rightarrow b\bar b)$. This background could be 
practically eliminated if both jets were required to be $b$-tagged. 
Presently, more data are being 
collected with an unprescaled $b$-tagged dijet trigger 
to yield a large sample of double-$b$-tagged dijet events with which to 
measure the rate for exclusive production in a low background 
environment. 

\subsubsection{Exclusive $\chi_c^0$ Production}
CDF has reported an upper limit of 
$49\pm 18\mbox{ (stat)}\pm 39\mbox{ (syst)}$~pb
for exclusive $\chi_c^0$ production
from a search for $J/\psi+\gamma$ events from 
$\bar pp\rightarrow \bar p+\chi_c^0(\rightarrow J/\psi+\gamma\rightarrow \mu\mu+\gamma)+\bar p$. 
Theoretical predictions of $\sim 70$~pb  
have recently been revised to $\sim 50$~pb~\cite{KMR,KKMRS}. More data, 
collected with a dedicated trigger, are currently being analyzed. 

\subsubsection{Exclusive $\gamma\gamma$ Production}
Data collected with a special trigger are currently being analyzed
in search for exclusive $\gamma\gamma$ production. The data at hand are 
already sufficient for placing an upper limit on the production cross section 
 at a level comparable to that of 
theoretical predictions~\cite{KMR,KKMRS}. With data continuing to come in, 
the process will hopefully be discoverd soon and be used, along with 
exclusive dijet production, for calibrating 
calculations for diffractive Higgs production at LHC. 

\subsection{Composite Pomeron?}
In the generalized renormalization model (see review in Ref.~\cite{lathuile}) 
the hard scattering in hard diffractive processes is  
controlled by the parton distribution function (PDF) at high $Q^2$,
while the diffractive rapidity gap is formed by the emission of a 
soft parton that neutralizes the color accross the rapidity region over 
which the hard exchange occurs. In this {\em composite Pomeron}\, model, 
the color neutralization does not have to extend over the entire rapidity 
region spanned by hard exchange, as is the case in models with 
particle-like Pomerons, {\it e.g.} in the BFKL model~\cite{BFKL}. 
Dijet events with a rapidity gap between jets offer an opportunity 
to differentiate between particle-like and composite 
Pomeron models.
 
\begin{figure}[h]
\psfig{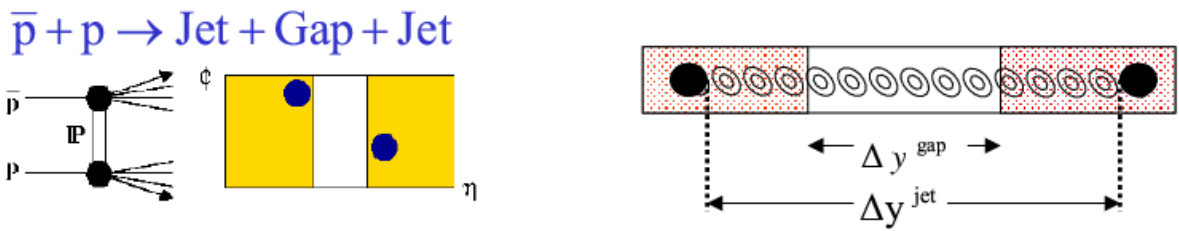}
\caption{Forward dijet production with a rapidity gap between jets: 
{(\em left)\,} for particle-like Pomerons, the gap extends all the 
way to the jets, while {(\em right)\,} for composite Pomerons, the gap 
width is generally smaller than the rapidity separation of the jets.}    
\label{fig:composite}
\end{figure}
 
Figure~\ref{fig:composite} shows schematic diagrams for the process 
$\bar p+p\rightarrow Jet+Gap+Jet$. For a particle-like Pomerons ({\em left}\,),
the gap extends all the way to the jets, while for composite Pomerons 
({\em right\,}) the gap is smaller than the rapidity separation of the jets.
At CDF, the cross section for events with jets in the forward Miniplug 
calorimeters  will be 
measured as a function of gap position and gap width. This information
will then be checked against predictions of various models 
to determine the nature of Pomeron exchange.   

\section{Conclusions}
A comprehensive set of studies of diffractive processes has been 
under way at CDF since the program was proposed in 1985, the year of the first 
{\em Bois Workshop} on Elastic and Diffractive Scattering. 
The results obtained on soft and hard diffraction in this twenty year 
period exhibit regularities that point to the QCD nature of the Pomeron.
The most striking feauture of the data is the $s$-independence of the 
single diffractive $d\sigma/dM^2$ differential cross section. 
This scaling behaviour provides a clue to understanding the  
Pomeron as a composite structure built up from a combination 
of gluons and/or quarks with the quantum numbers of the vacuum 
in accordance with QCD color constraints.    
Results from Run-II already reported at Conferences 
confirm Run-I results, data at hand are being analyzed to extend the 
Run-I findings, and studies are being carried out to provide information 
on rates and backgrounds relevant to formulating a strategic plan for 
carrying out the diffractive program proposed for the Large Hadron Collider.

\end{document}